\documentclass[conference]{IEEEtran}
\IEEEoverridecommandlockouts
\usepackage{cite}
\usepackage{amsmath,amssymb,amsfonts}
\usepackage{algorithmic}
\usepackage{graphicx}
\usepackage{textcomp}
\usepackage{xcolor}
\usepackage{multirow}
\usepackage{marginnote}
\PassOptionsToPackage{hyphens}{url}\usepackage{hyperref}

\begin{document}

\title{
RITA: Automatic Framework for Designing of Resilient IoT Applications \\
}

\author{\IEEEauthorblockN{1\textsuperscript{st} Luis Eduardo Pessoa}
 \IEEEauthorblockA{\textit{Federal University of Rio de Janeiro} \\
 Rio de Janeiro, Brazil \\
 lempb.pessoa@gmail.com}

 \and
 \IEEEauthorblockN{2\textsuperscript{nd} Cristóvão F. Iglesias Jr}
 \IEEEauthorblockA{\textit{University of Ottawa} \\
 Ottawa, Canada \\
 cfrei096@uottawa.ca}
 \and
 \IEEEauthorblockN{3\textsuperscript{rd} Claudio Miceli}
 \IEEEauthorblockA{\textit{Federal University of
 Rio de Janeiro} \\
 Rio de Janeiro, Brazil \\
 cmicelifarias@cos.ufrj.br}

}

\maketitle

\begin{abstract}
Designing resilient Internet of Things (IoT) systems requires i) identification of IoT Critical Objects (ICOs) such as services, devices, and resources, ii) threat analysis, and iii) mitigation strategy selection.
However, the traditional process for designing resilient IoT systems is still manual, leading to inefficiencies and increased risks. In addition, while tools such as ChatGPT could support this manual and highly error-prone process, their use raises concerns over data privacy, inconsistent outputs, and internet dependence. 
Therefore, we propose RITA, an automated, open-source framework that uses a fine-tuned RoBERTa-based Named Entity Recognition (NER) model to identify ICOs from IoT requirement documents, correlate threats, and recommend countermeasures.  RITA operates entirely offline and can be deployed on-site, safeguarding sensitive information and delivering consistent outputs that enhance standardization. 
In our empirical evaluation, RITA outperformed ChatGPT in four of seven ICO categories, particularly in actuator, sensor, network resource, and service identification, using both human-annotated and ChatGPT-generated test data. These findings indicate that RITA can improve resilient IoT design by effectively supporting key security operations, offering a practical solution for developing robust IoT architectures.
\end{abstract}


\section{Introduction}


Resilience should be addressed in the early stages of the development \cite{iglesias2023automated}. It  should be done in the design phase \cite{kossiakoff2020systems,kotonya1998requirements,herrmann2015engineering}. 
A general approach for designing a resilient IoT application during the design phase includes  three steps: 
(i) Identification of IoT critical objects (ICOs). This initial step focuses on identifying the essential elements within the IoT system that are vital for proper functionality and susceptible to IoT-specific threats. This is typically done by reviewing documents, such as storylines, user stories, and system requirements, to identify the key devices, resources, and services critical to the operation of the system \cite{delic2016resilience,bassi2013enabling,kossiakoff2020systems,herrmann2015engineering}. 
(ii) Threat Identification. Once critical components have been identified, the next step is to assess potential risks to the system. This includes identifying security, privacy, and reliability threats, which may arise from sources such as cyberattacks, malware, physical damage, or environmental conditions \cite{bassi2013enabling,kossiakoff2020systems,herrmann2015engineering}.
(iii) Selection of Mitigation Strategies. The final step is to specify effective measures to counteract the identified threats. Mitigation strategies may involve implementing security protocols, using encryption, adding redundancy, monitoring for anomalies, or applying physical security measures to safeguard critical IoT components \cite{bassi2013enabling,kossiakoff2020systems,herrmann2015engineering}.

Addressing resilience in the design phase allows \cite{kossiakoff2020systems,kotonya1998requirements,herrmann2015engineering,delic2016resilience,liu2010architectural}: (i) to deal with the complexity of the problems, (ii) effective communication, (iii) complete understanding, (iv) reduced costs, (v) predict behavior, (vi) reuse and (vii) analyze the feasibility (financial and practical) of the IoT system. 
However, designing resilient systems often requires a deep understanding of the potential sources of disturbances, their impacts, and how the system can respond to ensure continued operation. Furthermore, manually addressing resilience can be a significant challenge. As an effect, the manual process can be time-consuming and prone to error as it involves anticipating a wide range of potential scenarios and formulating appropriate responses.

\textbf{Motivation and problem specification.}
In \cite{iglesias2023automated} an investigation was made to determine the usefulness of NER models to identify IoT critical objects (services, devices, and resources) from IoT systems documents to sidestep the challenges of manual identification.
However, as described before, the identification of IoT critical objects is only the first step of the general procedures to design a resilient IoT application. Therefore, the following question remains open.
\textit{How can we automatically extract IoT critical objects from documents (storyline and requirements) and list all possible IoT threats and resilient countermeasures that can be used in the design of a resilient IoT application?}

\begin{figure}[htbp]
\centerline{\includegraphics[scale=0.5]{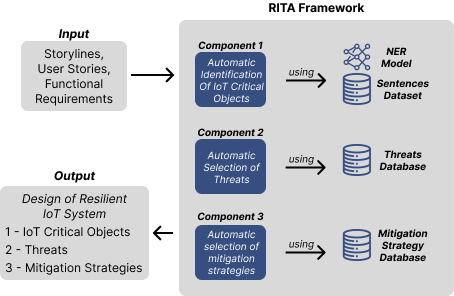}}
\caption{RITA Framework}
\label{fig-ria-architecture}
\end{figure}

\textbf{Our contributions.}
To address the problem above, we propose a framework named RITA. It  provides an environment for the automatic design of the \textbf{R}esilient \textbf{I}o\textbf{T} \textbf{A}pplication (RITA).
Three of the main components of the RITA can be seen in the Figure \ref{fig-ria-architecture} as well as the input and output. 
Component 1 of RITA addresses the automatic extraction of IoT critical objects with an NER model. It is a fine-tuned RoBERTa model, trained with the dataset created in this work.  The RoBERTa model is a BERT model with optimized hyperparameters \cite{liu2010architectural}. Upon processing a text input, the NER model will reply with the  identified  IoT Critical Object  and the category for the identified object from the text. It is done following the definitions from Table \ref{tab-ico-1}. 
Component 2 addresses the threat identification based on 
the relational nature of IoT critical objects and threats through an IoT threat database. 
Component 3 addresses the  selection of mitigation strategies.
It is also done based on a mitigation strategies database where the mitigation strategies related to the threats listed in component 2 are identified.
It is important to point out that the foundation for the three components of RITA framework is the architectural design decisions  for resilient IoT (ADD4RIOT) \cite{iglesias2023architectural}.  The NER model, IoT threats, and mitigation strategies database follow the concepts defined in ADD4RIOT.
It is a meta-model used to guide the design of resilient IoT applications. It systematically defines resilient IoT applications and their requirements: monitoring, detection, protection, restoration, and memorization. 


Therefore, the main contributions of this work are the following. 
 i) A large-sized datasets with annotations regarding the niche domain of IoT critical objects, with 66108 annotated phrases, and 1813 examples of IoT Critical Objects. 
 ii) A relational database representing the taxonomy defined in \cite{iglesias2023architectural} that can be used to automate the process of threats and mitigations identification. The database contains 97 countermeasure examples and 138 registered threats, all related to the 7 IoT Critical Objects categories listed in this work, which can also be seen in Figure \ref{fig-dataset-distribution}. iii) A trained RoBERTa model that is capable of identifying IoT critical objects in texts pertaining to the IoT system development domain, and
 iv) RITA framework which has a performace similiar to GPT-3 Model (baseline) on the task of IoT Critical Objects identification in 4 categories out of the 7 proposed for the NER task in this study, that is the result of the integration of i), ii) and iii). All source codes and the dataset used in the empirical evaluation are available in the GitHub repository 
 \footnote{ (\url{https://github.com/LEpessoa/RITA})}.
In addition, the RITA framework is also available for download and local use in the same repository $^1$.

\textbf{Advantages.} The advantages of using the RITA framework are i) the use of an offline tool that can be used without concerns for data breaches. ii) No internet dependance and on premise deployment capabilities. ii) All code is opensource making it a tool where you can understand what you run. iii) The output is predictable, as opposed to chatGPT where equal inputs can result in different outputs. This is an important characteristic of such a model, since the standardization of the identification of IoT critical objects is paramount to the development of resilient architectures. The intrinsic unreliability of chatGPT answers make it prone to the same false positive and false negative mistakes of manual identification.  iv) No limit to the number of prompts analyzed, as opposed to an hourly limit in off the shelf options such as chatGPT.

\begin{table}[htbp]
\caption{IoT Critical Objects (ICOs) Definition}
\begin{center}
\begin{tabular}{ p{0.2cm}|p{1cm}|p{3.4cm}|p{3.7cm}  }
            \hline
            \multicolumn{2}{c|}{\textbf{ICO}} & \textbf{Definition} &\textbf{Example}\\
            \hline
            \multirow{4}{4em}{\rotatebox[origin=c]{90}{Device}}
            &Sensor& Device that gathers information from real world entities.& Temperature, Humidity, Pressure, Light, Proximity, Motion, Accelerometers, Gyroscopes, Magnetometers\\
            \cline{2-4}
            &Actuator&  Device that makes a change in the physical state of one or more physical entities.&  Servo Motors, Linear Actuators, Solenoids, Valves, Relays, LED Lights, Heating Elements\\
            \cline{2-4}
            &Tag& Labels attached to physical entities with the aim of tracking position, existence or information regarding the physical entity& RFID (Radio Frequency Identification), NFC (Near Field Communication), Bluetooth Low Energy (BLE), GPS (Global Positioning System), QR Code (Quick Response), Barcode\\
            \cline{2-4}
            &Smart Camera& Advanced embedded systems that integrating image capturing capabilities with processing and networking to achieve varied purposes.&Facial recognition, license plate recognition, detection of defects or anomalies, Surveillance and security, fall detection, residential security and surveillance.\\
            \hline
            \multirow{4}{*}{\rotatebox[origin=c]{90}{Resource}}
            &Network Resource& Resources accessible via the network& Data repositories, ERP da-
tabases, cloud storage\\
            \cline{2-4}
            &On Device Resource& Software accessible without need of network, its accessible by the device itself and installed into it.&Device driver, Data cache,dados em RFID tags, Pro-gramming API\\
            \hline
            \multirow{4}{*}{\rotatebox[origin=c]{90}{Service}}
            &Service&An abstraction over resources and virtual entities, providing access to functionalities and information about the associated virtual entity or resource.&Web service or Local service, expressed in the form of an imperative verb, like Alert, Monitor, and Connect\\
            \hline
        \end{tabular}
\label{tab-ico-1}
\end{center}
\end{table}

\section{Related work}
Natural language processing techniques have been used by different authors with different objectives in the field of Requirements Engineering (RE), such as classification of requirements into functional/non-functional, classification of online product review, detection of redundant requirements and information extraction by leveraging NER models \cite{nadeau2007survey}. There also has been a research about the development of models to capture key elements of natural language requirements \cite{malik2021named,veera2019nerse}. In \cite{robeer2016automated}, it is proposed a pipeline for the automated creation of conceptual models from user stories with the application of natural language heuristics. Similar work was proposed in \cite{elallaoui2018automatic}. Existing NER models are typically trained in general categories, such as PERSON, LOCATION and ORGANIZATION, which are not useful in the niche paradigm of the IoT Systems\cite{malik2021named,robeer2016automated,veera2019nerse,elallaoui2018automatic}. As an extension of the categories defined in \cite{iglesias2023automated}, we have defined 7 subcategories to further subdivide Devices, Resources and Services. These categories can be seen on Table \ref{tab-ico-1} where the first column has the name of the categories defined in \cite{iglesias2023automated} and the sub categories are nested on the second column. As can be seen, Devices were subdivided into Actuators, Sensors, Tags and Smart Cameras. Resources were subdivided into Network Resources and On-Device Resources, and Services were not subdivided. Our work is the first to report the extraction of IoT Critical Objects with this fine grained categorization.

\section{Technical Approach}
\textit{Named Entity Recognition (NER)} is a Natural Language Processing task where entities are identified in text. The categories here defined were Actuators, Tags, Sensors, Smart Cameras, On-Device Resources, Network Resources and Services. The definition for each category can be seen on Table \ref{tab-ico-1}. For the identification process of the specific niche categories of the IoT Domain, we adopted the best architecture, BERT \cite{iglesias2023automated}. The model chosen for our NER task was the RoBERTa model available through the Spacy framework \footnote{(\url{https://spacy.io/models})}. The RoBERTa model has the same architecture as the BERT model, performing better due to the pre training approach adjustments that involve the time taken to train the model, the size of the batches, size of sequences and dynamically changing the masking pattern applied to the training data \cite{liu2019roberta}. The fine tuning of the RoBERTa model was done with a regular split of the dataset with 30\% for test and 70\% for trainig. For the evaluation of GPT-3, a group of 300 phrases were chosen, where 100 of these phrases represented Storylines phrases. 100 represented User Stories, and another 100 represented requirements. The storyline phrases were taken from the test group of the dataset. The other 200 phrases were generated using GPT-3. This group of 300 phrases formed the validation set (Table \ref{fscore-full-table} third column). The input of the model is some text representing a user story, storyline or requirement such as \textit{"The A3144E Hall Effect Sensor Switch is used in the smart garage door opener, which allows you to open and close the garage door remotely"}. The output would be \textit{("a3144e hall effect sensor switch","ACTUATOR")}

\section{Empirical Evaluation}
Our evaluation aims to answer the following Research Question (RQ): 
1) What is the performance of RITA and ChatGPT to identify IoT Critical Objects?

\textbf{Description of The Dataset:} It was not possible to find an annotated dataset with categories pertaining to the niche domain of IoT systems besides the dataset described in \cite{iglesias2023automated}. We build over the previous dataset of \cite{iglesias2023automated}, with the addition of 7121 phrases. Moreover, the original dataset was sub categorized to split 4 categories of the original dataset into the 7 categories found in the current dataset and create a large-sized dataset \cite{gemkow2018automatic-large-dataset}. The original dataset was generated utilizing two approaches.

 For the first approach, to annotate sentences for devices and resources, online sentence dictionaries
 \footnote{(\url{https://www.wordhippo.com/})}
 \footnote{(\url{https://sentence.yourdictionary.com/})}
 were  used, and the phrases were found by searching for examples of the proposed entities derived from their definitions in Table \ref{tab-ico-1}. The phrases found in the online dictionaries were annotade using Doccano
 \footnote{(\url{https://github.com/doccano/doccano})}, an open source text annotation tool. The second approach was used for the services category. The CrowdRe
 \footnote{(\url{https://crowdre.github.io/murukannaiah-smarthome-requirements-dataset})} dataset was used, consisting of 2966 requirements sentences in the form of user stories. The phrases in this dataset are well defined, separated into role, feature and benefit. Some of the verbs inside feature fitted the description for service and thus were annotated using NLTK in python\cite{iglesias2023automated}.
The phrases generated in this work were created using a third approach, with a well defined query pipeline. First for a given category, for example sensors, we queried chatGPT to find examples of sensors, such as light, temperature and gas sensors. Then for temperature sensor, we asked for commercially available options of this given type of sensor. And last, with each commercially available type of sensor we asked for the phrases.
\footnote{Process Example Online: (\url{https://chatgpt.com/share/671aa58d-083c-8005-8965-c848fac2b7de})}
The phrases were fed to a program that checked for the presence of the target word and wrote the formatted data point to a CSV file on the local disk. The program consisted of a terminal only interface that generated the prompt asking for the phrases and then awaited for the user to give the generated phrases back for verification. With this approach there is less leeway for human error.

\begin{figure}[htbp]
\hspace{-0.8cm}
\centerline{\includegraphics[scale=0.45]{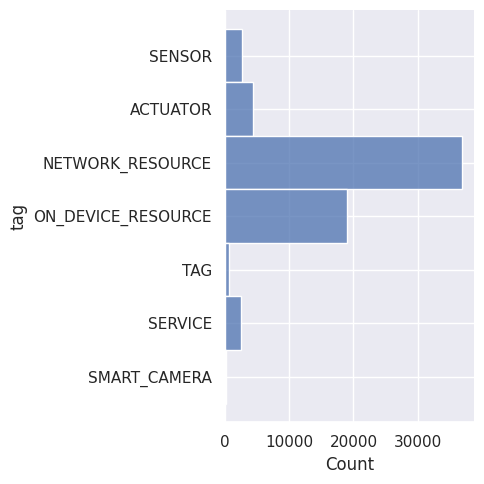}}
\caption{Dataset Distribution}
\label{fig-dataset-distribution}
\end{figure}


\textbf{Results and Discussion:} To compare the RITA Framework's performance with chatGPT, specific prompts were made so that chatGPT could label the phrases with the same format the trained NER model does, aiming to ensure a fair comparison. From the test portion of the dataset, 300 phrases were chosen randomly. A Few Shot prompting approach was used to get the answers from chatPGT. The few-Shot approach proved to be the one yielding the best results in terms of consistency of the replies and maintenance of the format suggested for the answers requested from chatGPT. The format suggested consisted of a parenthesis with two strings, one for the entity matched, and another for the category. This format is consistent with the format that a program could have as output, using the NER model. The format constraint also helps to ensure a fair comparison where both chatGPT and the  NER model could be used interchangeably in an automated pipeline. Given the answer provided, if the format was respected, we proceed to classify the answer as a true positive or a false positive. If the response predicted the right category and had overlap it's deemed a true positive, otherwise it's a false positive. False negatives were seen in occasions where chatGPT reported to not have found any IoT Critical Object in phrases that did contain them. The RITA framework reported no false negatives. Also, even with the format constraint, chatGPT should have the upper hand since many of the phrases were generated using chatGPT itself. In Table \ref{fscore-full-table}, on the Test column, The F-Score reported from the evaluation of the RITA Framework on the testing set (second column of Table \ref{fscore-full-table}), shows that the framework can accurately identify ICOs on not before seen phrases. The F-Score is a good measure for the evaluation of the RITA Framework given the unbalanced characteristic of our dataset, as can be seen on Figure \ref{fig-dataset-distribution}. The F-Score is the harmonic mean of the Precision and recall calculated from the true positives, false positives and false negatives, giving an overall metric to evaluate the model. Since the dataset has 7 entities, The F-Score is calculated for each one of them. The F-Score helps mitigating the imbalance of the dataset, being the most common metric employed in this kind of problem \cite{brownlee2020imbalanced}. The imbalance characteristic was a natural outcome of the occurrence of the different types of IoT Critical Objects found in texts in the web.
The results regarding chatGPT's comparison to the NER model can be seen on Table \ref{fscore-full-table}, in the third column, that shows the result of the comparison on the validation set. The comparison results show that the RITA Framework bests chatGPT on 4 of the 7 ICO categories, even though the phrases that chatGPT evaluated were created using chatGPT. The categories where the RITA Framework performed better were actuator, sensor, network resource and service, while faring worse in the categories tag, smart camera and on-device resource.



\begin{table}[h]
	\begin{center}
	  \caption{F-Score Comparison}
    \label{fscore-full-table}
		\begin{tabular}{c|c|ccc}\hline
		   & \textbf{Testing set} & \multicolumn{2}{|c}{\textbf{Validation set}}\\ \hline 
              & \textbf{RITA} & \textbf{RITA} & \textbf{ChatGPT}\\ 
            \textbf{Category} & \textbf{(RoBERTa)} & \textbf{(RoBERTa)} & \textbf{(GPT-3)}\\\hline
		  Actuator & 0.9740831296 & 0.83 & 0.71 \\ \hline
        Tag & 0.9512195122 & 0.84 & 0.96 \\ \hline
        Sensor & 0.9740831296 & 1.00 & 0.96 \\ \hline
        Smart Camera & 0.9370629371 & 0.76 & 0.94 \\ \hline
        On-Device Resource & 0.9967506806 & 0.67 & 0.80 \\ \hline
        Network Resource & 0.9982910595 & 0.74 & 0.59 \\ \hline 
        Service & 0.8931830381 & 0.60 & 0.47 \\ \hline
		\end{tabular}
	\end{center}	
\end{table}

\section{Conclusion}

This work presents RITA, an automated framework for identifying IoT Critical Objects (ICOs) essential to resilient IoT applications. By leveraging a fine-tuned RoBERTa model, RITA consistently demonstrated competitive performance, surpassing ChatGPT in key ICO categories, including actuators, sensors, network resources, and services, while addressing significant limitations in data privacy, internet dependency, and output consistency associated with online tools. Our empirical evaluations affirm that RITA is a viable, secure alternative for resilience-focused IoT development, capable of functioning entirely offline and mitigating risks inherent to sensitive data processing in IoT environments.
Future research could enhance RITA’s adaptability across all ICO categories, improving its performance in areas like smart cameras and tags. Additionally, scaling capabilities of the RITA to accommodate broader IoT architectures will enable its application in increasingly complex environments. The RITA framework marks a step forward in resilient IoT system design, supporting automated, consistent, and secure handling of critical security functions.



\bibliographystyle{IEEEtran}
\bibliography{conference_101719}

\begin{thebibliography}{10}
\providecommand{\url}[1]{#1}
\csname url@samestyle\endcsname
\providecommand{\newblock}{\relax}
\providecommand{\bibinfo}[2]{#2}
\providecommand{\BIBentrySTDinterwordspacing}{\spaceskip=0pt\relax}
\providecommand{\BIBentryALTinterwordstretchfactor}{4}
\providecommand{\BIBentryALTinterwordspacing}{\spaceskip=\fontdimen2\font plus
\BIBentryALTinterwordstretchfactor\fontdimen3\font minus \fontdimen4\font\relax}
\providecommand{\BIBforeignlanguage}[2]{{%
\expandafter\ifx\csname l@#1\endcsname\relax
\typeout{** WARNING: IEEEtran.bst: No hyphenation pattern has been}%
\typeout{** loaded for the language `#1'. Using the pattern for}%
\typeout{** the default language instead.}%
\else
\language=\csname l@#1\endcsname
\fi
#2}}
\providecommand{\BIBdecl}{\relax}
\BIBdecl

\bibitem{iglesias2023automated}
C.~F. Iglesias, R.~Guo, P.~Nucci, C.~Miceli, and M.~Bolic, ``Automated extraction of iot critical objects from iot storylines, requirements and user stories via nlp,'' in \emph{2023 10th IEEE Swiss Conference on Data Science (SDS)}.\hskip 1em plus 0.5em minus 0.4em\relax IEEE, 2023, pp. 104--107.

\bibitem{kossiakoff2020systems}
A.~Kossiakoff, S.~M. Biemer, S.~J. Seymour, and D.~A. Flanigan, \emph{Systems engineering principles and practice}.\hskip 1em plus 0.5em minus 0.4em\relax John Wiley \& Sons, 2020.

\bibitem{kotonya1998requirements}
G.~Kotonya and I.~Sommerville, \emph{Requirements engineering: processes and techniques}.\hskip 1em plus 0.5em minus 0.4em\relax Wiley Publishing, 1998.

\bibitem{herrmann2015engineering}
J.~W. Herrmann, \emph{Engineering decision making and risk management}.\hskip 1em plus 0.5em minus 0.4em\relax John Wiley \& Sons, 2015.

\bibitem{delic2016resilience}
K.~A. Delic, ``On resilience of iot systems: The internet of things (ubiquity symposium),'' \emph{Ubiquity}, vol. 2016, no. February, pp. 1--7, 2016.

\bibitem{bassi2013enabling}
A.~Bassi, M.~Bauer, M.~Fiedler, T.~Kramp, R.~Van~Kranenburg, S.~Lange, and S.~Meissner, \emph{Enabling things to talk}.\hskip 1em plus 0.5em minus 0.4em\relax Springer Nature, 2013.

\bibitem{liu2010architectural}
D.~Liu, R.~Deters, and W.-J. Zhang, ``Architectural design for resilience,'' \emph{Enterprise Information Systems}, vol.~4, no.~2, pp. 137--152, 2010.

\bibitem{iglesias2023architectural}
C.~F. Iglesias~Jr, C.~Miceli, and M.~Bolic, ``An architectural design decision model for resilient iot application,'' \emph{arXiv preprint arXiv:2306.10429}, 2023.

\bibitem{nadeau2007survey}
D.~Nadeau and S.~Sekine, ``A survey of named entity recognition and classification,'' \emph{Lingvisticae Investigationes}, vol.~30, no.~1, pp. 3--26, 2007.

\bibitem{malik2021named}
G.~Malik, M.~Cevik, Y.~Khedr, D.~Parikh, and A.~Basar, ``Named entity recognition on software requirements specification documents.'' in \emph{Canadian Conference on AI}, 2021.

\bibitem{veera2019nerse}
M.~Veera Prathap~Reddy, P.~Prasad, M.~Chikkamath, and S.~Mandadi, ``Nerse: named entity recognition in software engineering as a service,'' in \emph{Service Research and Innovation: 7th Australian Symposium, ASSRI 2018, Sydney, NSW, Australia, September 6, 2018, and Wollongong, NSW, Australia, December 14, 2018, Revised Selected Papers 7}.\hskip 1em plus 0.5em minus 0.4em\relax Springer, 2019, pp. 65--80.

\bibitem{robeer2016automated}
M.~Robeer, G.~Lucassen, J.~M.~E. Van Der~Werf, F.~Dalpiaz, and S.~Brinkkemper, ``Automated extraction of conceptual models from user stories via nlp,'' in \emph{2016 IEEE 24th international requirements engineering conference (RE)}.\hskip 1em plus 0.5em minus 0.4em\relax IEEE, 2016, pp. 196--205.

\bibitem{elallaoui2018automatic}
M.~Elallaoui, K.~Nafil, and R.~Touahni, ``Automatic transformation of user stories into uml use case diagrams using nlp techniques,'' \emph{Procedia computer science}, vol. 130, pp. 42--49, 2018.

\bibitem{liu2019roberta}
Y.~Liu, M.~Ott, N.~Goyal, J.~Du, M.~Joshi, D.~Chen, O.~Levy, M.~Lewis, L.~Zettlemoyer, and V.~Stoyanov, ``Roberta: A robustly optimized bert pretraining approach,'' \emph{arXiv preprint arXiv:1907.11692}, 2019.

\bibitem{gemkow2018automatic-large-dataset}
T.~Gemkow, M.~Conzelmann, K.~Hartig, and A.~Vogelsang, ``Automatic glossary term extraction from large-scale requirements specifications,'' in \emph{2018 IEEE 26th International Requirements Engineering Conference (RE)}.\hskip 1em plus 0.5em minus 0.4em\relax IEEE, 2018, pp. 412--417.

\bibitem{brownlee2020imbalanced}
J.~Brownlee, \emph{Imbalanced classification with Python: better metrics, balance skewed classes, cost-sensitive learning}.\hskip 1em plus 0.5em minus 0.4em\relax Machine Learning Mastery, 2020.

\end{thebibliography}
\end{document}